\documentclass[english]{article}
\usepackage[T1]{fontenc}
\usepackage[latin9]{inputenc}
\usepackage{geometry,amsmath}
\usepackage{amssymb} 
\usepackage{amsmath} 
\usepackage{physics}
\usepackage{color,graphicx}
\usepackage{array}
\date{}
\makeatletter

\usepackage{cite,times}

\makeatother

\usepackage{babel}
\begin{document}

\title{Continuous variable controlled quantum dialogue and secure multiparty quantum computation}

\author{Ashwin Saxena$^{1,}$\thanks{ashn5.new@gmail.com}, Kishore Thapliyal$^{1,2,}$\thanks{tkishore36@yahoo.com}, Anirban Pathak$^{1,}$\thanks{anirban.pathak@gmail.com}\\
$^{1}$Jaypee Institute of Information Technology, A-10, Sector-62,
Noida, UP-201307, India~\\
$^{2}$RCPTM, Joint Laboratory of Optics of Palacky University
and Institute of Physics~\\ of Academy of Science of the Czech Republic,
Faculty of Science, Palacky University, ~\\ 17. listopadu 12, 771 46 Olomouc,
Czech Republic}

\maketitle

\begin{abstract}
A continuous variable controlled quantum dialogue scheme is proposed. The scheme is further modified to obtain two other protocols of continuous variable secure multiparty computation. The first one of these protocols provides a solution of two party socialist millionaire problem, while the second protocol provides a solution for a special type of multi-party socialist millionaire problem which can be  viewed as a protocol for multiparty quantum private comparison. It is shown that the proposed scheme of continuous variable controlled quantum dialogue can be performed using bipartite entanglement and can be reduced to obtain several other two and three party cryptographic schemes in the limiting cases. The security of the proposed scheme and its advantage over corresponding discrete variable counterpart are also discussed. Specifically, the ignorance of an eavesdropper in the proposed scheme is shown to be very high compared with corresponding discrete variable scheme and thus the present scheme is less prone to information leakage inherent with the discrete variable quantum dialogue based schemes.It is further established that the proposed scheme can be viewed as a continuous variable counterpart of quantum cryptographic switch which allows a supervisor to control the information transferred between the two legitimate parties to a continuously varying degree.
\end{abstract}

\section{Introduction}

With RSA \cite{rivest1978method} and similar classical cryptographic schemes \cite{diffie1976new} facing attacks from the advent of scalable quantum computers in the near future, quantum cryptography provides a necessary alternative. As the security in secure quantum communication, unlike classical cryptography, is not conditioned upon any assumption regarding computational powers of an eavesdropper and comes from the physical laws described by quantum mechanics, it is expected to provide unconditional security (see \cite{gisin2002quantum,pathak2013elements} for review). At times this unconditional security is analyzed either within the domain of quantum mechanics \cite{acin2006bell} or post-quantum theories \cite{aravinda2015orthogonal} as well as exploiting imperfections in the devices used to implement certain quantum cryptographic scheme (see \cite{lo2014secure} for review). 

The first quantum cryptographic scheme which allowed two distant parties Alice and Bob to procure secure random key using quantum resources is now known as BB84 protocol for quantum key distribution (QKD) \cite{bennett1984quantum}. This pioneering work was followed by several QKD schemes \cite{ekert1991quantum,goldenberg1995quantum,PhysRevLett.99.140501,lo2012measurement} and the schemes for several other quantum cryptographic tasks \cite{long2002theoretically,bostrom2002deterministic,lucamarini2005secure,thapliyal2015applications,sharma2016comparative,banerjee2016asymmetric,thapliyal2017quantum,shukla2017semi,banerjee2018quantum}.
Among these numerous quantum cryptography schemes, direct secure quantum communication schemes are interesting as they allow to perform secure communication without prior  generation/distribution of key \cite{long2002theoretically,bostrom2002deterministic,lucamarini2005secure},  and thus it resolves all the issues with secure key distribution and key management. The direct communication schemes are accomplished by splitting the useful information after encoding secret message in such a manner that the sender (Alice) reveals a part of this to the receiver (Bob) to decode her message only after confirmation of no eavesdropping attempt. In the meanwhile, Eve's attacks fail to reveal her anything in the absence of Alice's part of the information. Generalizations of these direct secure quantum communication schemes for two-way \cite{nguyen2004quantum,banerjee2016asymmetric,shukla2017semi}, three-party controlled \cite{thapliyal2015applications,shukla2017semi}, and multiparty \cite{banerjee2018quantum} communication tasks are also proposed in the past. Information is encoded on discrete systems in all these quantum communication schemes, and photon counters are used to decode the secret. In contrast, continuous variable (CV) quantum cryptography schemes for QKD are also proposed in the past \cite{ralph1999continuous,hillery2000quantum,reid2000quantum,ralph2000security,PhysRevA.63.022309, grosshans2002continuous} where information is encoded by modulating the quadratures and decoded by homodyne or heterodyne measurement.
Recently, it has been shown that CV systems provide better  key generation rate and performance than that obtained in  discrete variable (DV) QKD systems for relatively high transmittance channel \cite{lasota2017robustness}. CV schemes allow to encode a large amount of information which can also be used in broadband  transmission. Further, the implementation of CV schemes requires relatively inexpensive devices as in CV schemes, one can use commercially available room temperature highly efficient and large bandwidth homodyne detectors and the existing optical communication technology, but in  a DV scheme, one would require cryogenic single photon detectors  and efficient single photon source \cite{pirandola2015reply}. Thus, CV schemes fulfill the requirements for realistic metropolitan networks  \cite{pirandola2015reply}. Therefore, using the pre-functioning architecture in place \cite{fossier2009field}, quantum tasks with CV systems are expected to find a place in the future secure communication technology \cite{pirandola2015reply}.  

CV counterparts of the direct secure quantum communication protocols are also proposed  \cite{pirandola2008quantum,yu2016efficient,gong2017multiparty,gong2018continuous,gong2018quantum,yuan2015continuous,marino2006deterministic} in the recent past. Specifically, a protocol for CV quantum dialogue (QD) was proposed \cite{zhou2017new} using two-mode  squeezed state as resource and CV Bell measurement performed with homodyne detectors. A natural extension to this scheme is  to carry out this QD task in a manner such that a controller governs/supervises the flow of information between the two legitimate parties, known as controlled QD (CQD) scheme. The controller has the power to terminate the proceeding if he wishes or comes across a malfunctioning in the procedure. In case of DV CQD \cite{dong2008controlled,thapliyal2015applications}, Charlie withholds either information regarding channel preparation (i.e., Bell state shared between Alice and Bob) or final measurement outcome which restricts Alice and Bob to gain the information of each other. More recently, it has been shown that several other two and three party quantum cryptographic schemes can be reduced from a secure CQD scheme \cite{thapliyal2017quantum}.

A closely associated field of research is secure multiparty computation \cite{yao1982protocols}, i.e., to compute a function  with inputs given by more than one party in a secure manner. Recently, some of the present authors have shown that some direct secure quantum communication schemes can be used as primitives of quantum solutions for several socioeconomic problems, which can be introduced as secure multiparty computation tasks \cite{shukla2017semi,thapliyal2016protocols,sharma2017quantumauction,thapliyal2018orthogonal}. Motivated by the fact that there is no CV counterpart of CQD scheme, which can be used as primitive for quantum voting \cite{thapliyal2016protocols}, quantum e-commerce \cite{shukla2017semi,thapliyal2018quantum}, and quantum private comparison \cite{thapliyal2018orthogonal}, here we have proposed a CV CQD scheme and shown that it can be used to provide  the solutions for the socialist millionaire problem (SMP) in two party case and a solution for a particular type of multiparty SMP, known as multiparty quantum private comparison. 

The rest of the paper is organized as follows. In Section \ref{sec:res}, we briefly discuss generation of quantum state, two-mode squeezed state, and dense coding used in our CQD scheme. Subsequently, a CV CQD scheme is proposed in Section \ref{sec:prot}. The application of the proposed scheme to solve  SMP for two parties with an extension to carry out this task for the case of $n$-parties are reported in Sections \ref{sec:SMP}. Finally, we discuss the security of the proposed scheme in Section \ref{sec:sec} before concluding in Section \ref{sec:con}.

\section{Two-mode squeezed state generation and properties\label{sec:res}}

    There are various methods for generating two-mode squeezed states \cite{ya2015compact,zhang2000experimental,laurat2005entanglement}. We briefly discuss the process of generating such a state using nondegenerate optical parametric amplification (NOPA) process, which employs the nonlinear optical parametric down conversion process to produce entangled beams in orthogonal polarization \cite{andersen201630,jia2004experimental}. Specifically, in NOPA, we may initially start with a Nd:YAP/KTP laser to generate fundamental and second-harmonic waves as the seed and pump field inputs for NOPA. Subsequently, these waves are sent through a type-II nonlinear KTP crystal to produce a pair of entangled light beams through the parametric down conversion process (see \cite{jia2004experimental} for experimental implementation). This process of producing a two-mode entangled squeezed state is equivalent to theoretically applying a two-mode squeezing operator, defined as
\begin{equation}\label{eq: 1}
S (r ) = e^{ir(a^{\dagger}_{11}a^{\dagger}_{12} - a_{11}a_{12})  },
\end{equation}
where $r$ is squeezing parameter.

Using the two-mode squeezed state, CV information can be transmitted using dense coding with the help of Bell measurement \cite{ya2015compact}, which reveals the difference of amplitude quadratures and the sum of the phase quadratures of the input modes. The dense coding capacity for a two-mode squeezed state has been evaluated \cite{braunstein2000dense} to be equivalent to $C = \ln(1+\bar{n} +\bar{n}^{2})$, where $\bar{n}$ is the average number of  photons defined as $\bar{n}=\sigma^{2} +\sinh^{2}(r)$,  with $\sigma$ as the variance in the probability distribution obtained after homodyne detection.

\begin{center}
\begin{figure}[h]
\centering
 \includegraphics[width=\linewidth]{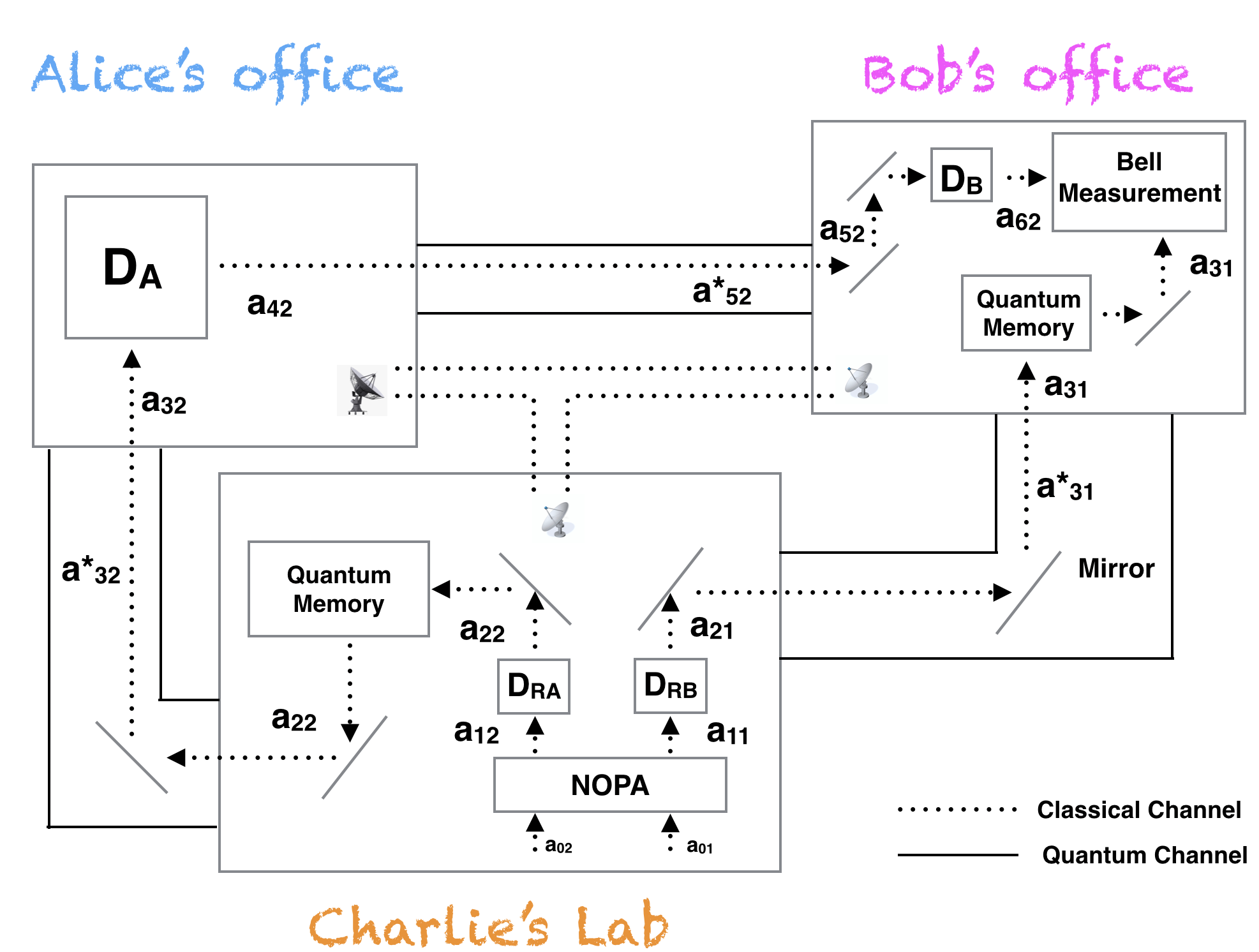}
\caption{\label{fig:scheme}(Color online) A schematic diagram for implementation of CV CQD.} 
\end{figure}
\end{center}

\section{Controlled quantum dialogue\label{sec:prot}}
In this section, we present our CQD scheme between two spatially separated parties Alice and Bob under the supervision of Charlie. The protocol is defined in the following steps.
\begin{description}
\item{Step~1}. Charlie displaces the two input vacuum modes $a'_{01}$ and $a'_{02}$ with $D (\alpha = x^{\prime} + ix^{\prime})$ and $D (\beta = y^{\prime} +iy^{\prime})$ to obtain two new modes $a_{01} = D^{\dagger} (\beta) a'_{01} D (\beta) $ and $ a_{02} = D^{\dagger}(\beta) a'_{02} D (\beta)$, respectively. Thereafter, $a_{01}$ and $a_{02}$ are the input modes of NOPA. The evolution of these modes can be described by application of $S(r)$, given in Eq. \eqref{eq: 1} on $a_{01}$ and $a_{02}$ to generate the two-mode squeezed vacuum states $a_{11}$ and $a_{12}$, with amplitude and phase given by  
\begin{equation}
\begin{split}
&\ X_{a_{11}} = X_{a_{01}}\cosh(r) \pm X_{a_{02}} \sinh(r),\\
&\ P_{a_{11}}  =P_{a_{01}} \cosh(r)  \pm P_{a_{02}}  \sinh(r),\\
&\ X_{a_{12}} = X_{a_{02}} \cosh(r) \pm X_{a_{01}} \sinh(r),\\
&\ P_{a_{12}}  = P_{a_{02}} \cosh(r) \pm P_{a_{01}} \sinh(r).\\
\end{split}
\end{equation}
Subsequently, Charlie performs Bell Measurement on the output modes to obtain
\begin{equation}\label{eq: 2}
\begin{split}
&\ X_{\mu_{0}}  = (X_{a_{11}} - X_{a_{12}})/\sqrt{2},\\
&\ P_{\mu_{0}}  =  (P_{a_{11}} + P_{a_{12}})/\sqrt{2}\\
\end{split}
\end{equation}
and broadcasts this information publicly.

\item{Step~2}. Charlie uses the same two input vacuum modes $a'_{01}$, $a'_{02}$, and the same two-mode squeezing operation $S(r)$ to produce the two-mode squeezed vacuum states  $a_{11}$ and $a_{12}$ as in Step 1. The schematic diagram is shown in Fig. \ref{fig:scheme}. Charlie creates two random sequences $R_{A}$ and $R_{B}$ with $R_{i_{j}}\in\mathbb{C}$   for displacing the two modes at various time frames $T_{i}$ in each mode to obtain $a_{21}$ and $a_{22}$ with
\begin{equation}
\begin{split}
&\ X_{a_{21}} = X_{a_{11}} + x_{R_{B}},\\
&\ X_{a_{22}} = X_{a_{12}} + x_{R_{A}},\\
\end{split}
\end{equation}  
and similarly for the phase quadratures of both modes.
 
\item{Step~3}. He sends the entangled optical mode $a_{21}$ to Bob via the quantum channel while he keeps the entangled optical mode $a_{22} $ with himself.

\item{Step~4}. On receiving the entangled optical mode $a^{*}_{31}$ (obtained after $a_{21}$ passes through the quantum channel with transmission efficiency $\eta$), Bob performs an amplification (defined by gain coefficient $g=1/\sqrt{\eta}$) on the mode and informs Charlie that he has received it. In fact, linear amplification is performed for all the transmitted modes and will not be discussed repeatedly. Bob and Charlie check for eavesdropping by measuring either the amplitude or phase in a given time frame $T_{1}$. If their measurement outcomes are found to be correlated,  they conclude that no eavesdropping is attempted.  Bob discards the pulses used for entanglement checking and stores the mode as  $a_{31}$.

\item{Step~5}. Charlie then sends the other entangled optical mode to Alice (via a quantum channel) who receives $a^{*}_{32}$. Alice performs linear amplification  on the mode, then Alice and Bob decide a time frame $T_{2}$ for checking eavesdropping. In this step, they can also verify entanglement in their modes and ask Charlie to disclose the value of his random displacement for the given time frame $T_{2}$. With the help of Charlie's announcement of his decoy value for the given time frame and their measurement outcomes, they can not only check eavesdropping, but can also detect a participant attack by Charlie as well (discussed in detail in Section \ref{sec:sec}). 

\item{Step~6}. Once the security of transmission is established, Alice encodes her information on mode $a_{32}$  in time frame $T_{4}$  by applying a displacement operation as $a_{42}  = D^{\dagger}(\alpha) a_{32} D (\alpha)$, where $ \alpha= (x_{A} + i p_{A})$ is the message she wants to send to Bob, leading to 
\begin{equation}
\begin{split}
&\ X_{a_{42}} = X_{a_{32}} + x_{A},\\
&\ P_{a_{42}} = P_{a_{32}} + p_{A}.\\
\end{split}
\end{equation}  
She also applies random displacement operation in time frame $T_{3}$ as decoy.

\item{Step~7}. Alice sends $a^{*}_{52}$ to Bob, who performs amplification on the mode and announces its receipt. Subsequently, he checks for eavesdropping with the help of information shared by Charlie and Alice regarding their random displacement operations corresponding to the time frame $T_{3}$. If there is eavesdropping they go to Step 2 else Bob displaces the mode $a_{52}$ in time frame $T_{4}$  with $D (\gamma)$, where $\gamma= (x_{B} + ip _{B})$ is the value that encodes the information that he wants to communicate, to obtain
\begin{equation}
\begin{split}
&\ X_{a_{62}} = X_{a_{52}} + x_{B},\\
&\ P_{a_{62}} = P_{a_{52}} + p_{B}.\\
\end{split}
\end{equation}  
\item{Step~8}. Bob then performs Bell measurement on both modes $a_{31}$ and $a_{62}$ in time frame $T_{4}$, to obtain 
\begin{equation}
\begin{split}
&\ X_{{\mu}_1} = (X_{a_{31}} - X_{a_{62}})/\sqrt{2}\\
&\ = (X_{a_{31}} - X_{a_{22}} - x_{B} - x_{A}  )/\sqrt{2}\\
&\ = (X_{a_{11}} - X_{a_{12}} - x_{B} - x_{A} + x_{R_{B_{T_{4}}}} - x_{R_{A_{T_{4}}}})/\sqrt{2} \\
&\ = (-x_{B} - x_{A} + \sqrt{2}(X_{\mu_{0}}) + x_{R_{B_{T_{4}}}} - x_{R_{A_{T_{4}}}})/\sqrt{2}.  \\
\end{split}
\end{equation}
Finally, Bob gets 
\begin{equation}
 X=\sqrt{2}(X_{\mu_{1}} - X_{\mu_{0}}) = x_{R_{B_{T_{4}}}} - x_{R_{A_{T_{4}}}} - x_{B} - x_{A} 
\end{equation}  
and
\begin{equation}
P=\sqrt{2}(P_{\mu_{1}} - P_{\mu_{0}}) =  p_{R_{B_{T_{4}}}} + p_{R_{A_{T_{4}}}} + p_{B} + p_{A}. 
\end{equation}
Hereafter, Bob announces his final measurements $X$ and $P$. 
At this step, Alice and Bob ask for Charlie's encoded information ($R_{i_{T_{4}}}$) in the time frame $T_{4}$, which Charlie reveals if he wants the dialogue task to be accomplished. Using Bob's measurement outcomes and Charlie's information, Alice obtains the information encoded by Bob and vice versa.
\end{description}
Here, it is also worth mentioning that Alice and Bob may either encode discrete $N$-bit messages by dividing the real number line into $2^N$ intervals \cite{zhou2017new} or continuous messages on both quadratures. This transmission of continuous messages may be useful in several other cryptographic schemes, where a prior shared CV secret is required (for instance, see\cite{darunkar2014braided,liu2017quantum}).
\begin{center}
\begin{figure}[t]
\centering
 \includegraphics[width=\linewidth]{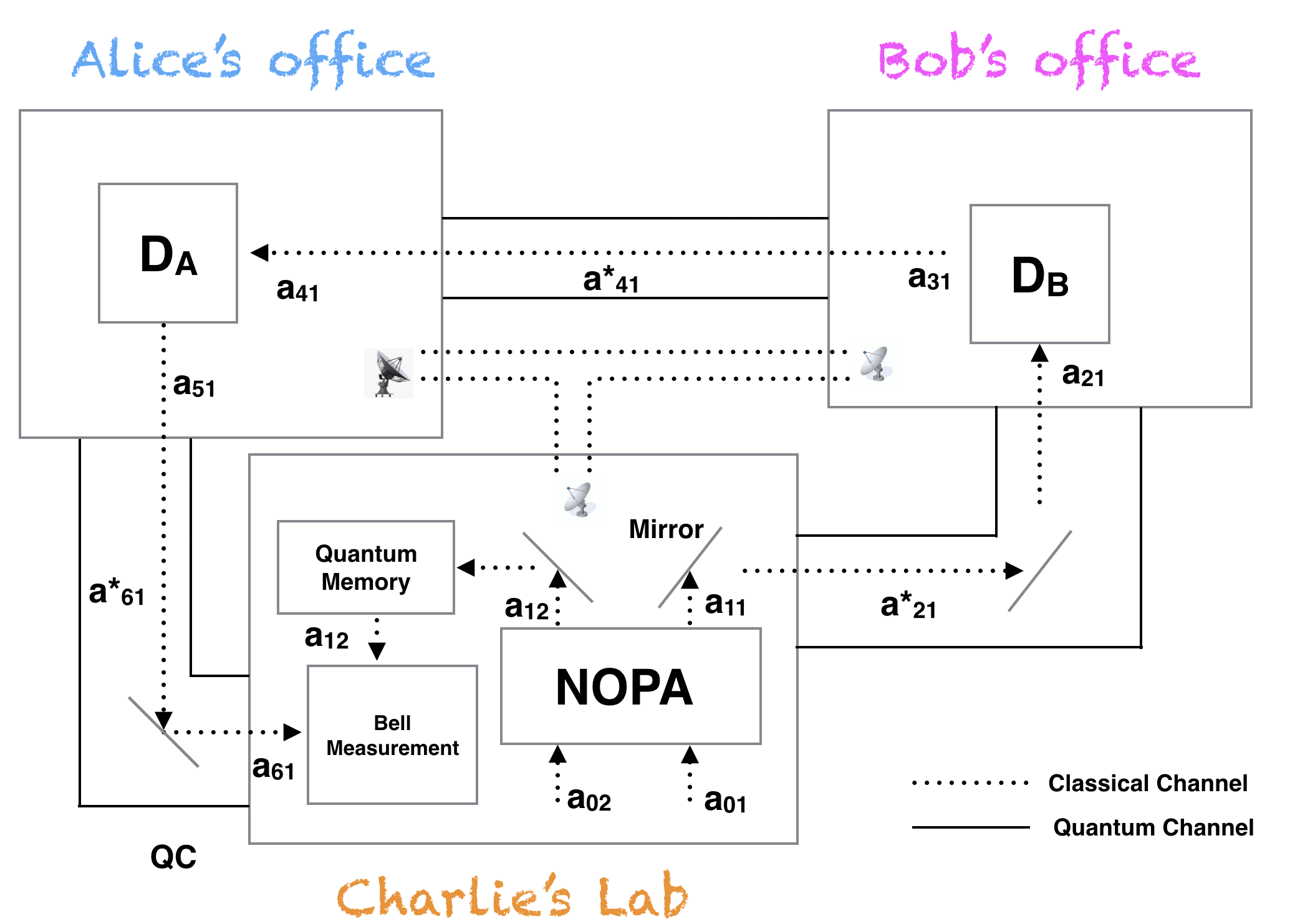}
\caption{\label{fig:SMP}(Color online) A schematic diagram for CV solution for two party SMP.}
\end{figure}
\end{center}

\section{Application: Socialist millionaire protocol\label{sec:SMP}}
Here we present a solution of SMP as an application of our CV CQD scheme. Firstly, we discuss a specific case in which two parties Alice and Bob wish to compare their assets with the help of an untrusted third party Charlie. In this task, the two parties (millionaires) wish to compare their assets (say $A$ and $B$ respectively), i.e., to compute whether $A>B$, $A<B$ or $A=B$ while maintaining  secrecy of $A$ and $B$ both from each other and an outsider.
\subsection{Two party socialist millionaire protocol\label{sec:SMP-2}}
The protocol (shown schematically in Fig. \ref{fig:SMP}) works as follows. 
\begin{description}
\item{Steps~1-4}. Same as Steps 1 to 4 of CQD in the previous section.

\item{Step~5}. Bob performs amplification and displaces the received mode : $a_{31}  = D^{\dagger}(\alpha) a_{21} D (\alpha) $, where $ \alpha= (x_{B} + ix_{B})$ is the amount of wealth $x_{B}$ (say in millions). Thus, the transformed quadrature can be written as 
\begin{equation}
X_{a_{31}} = X_{a_{21}} +x_{B}.
\end{equation}

\item{Step~6}. Bob sends $a_{31}$ to Alice, who receives $a_{41}$, performs amplification and informs him regarding it. Thereafter, they check for eavesdropping with the help of Charlie. If there is an eavesdropping attempt they abort and go back to Step 2 else Alice displaces the mode with $D (-\gamma)$ with $\gamma=(x_{A} + ix_{A})$ depending on the
amount of wealth $x_{A}$ (in millions). This leads to the following transformation
\begin{equation}
X_{a_{51}} = X_{a_{41}} - x_{A}.
\end{equation}  

\item{Step~7}. Alice sends $a_{51}$ to Charlie, who receives $a_{61}$. After  amplification and checking for eavesdropping, Charlie performs Bell measurement on $a_{12}$ and $a_{61}$, to obtain
\begin{equation}
\begin{split}
&\ X_{\mu_{1}} = (X_{a_{61}} - X_{a_{12}})/\sqrt{2},\\
&\ = (X_{a_{41}} - x_{A} - X_{a_{12}})/\sqrt{2},\\
&\ = (X_{a_{21}} + x_{B} - x_{A} - X_{a_{12}})/\sqrt{2},\\
&\ = (x_{B}- x_{A} - \sqrt{2}X_{\mu_{0}})/\sqrt{2}.\\
\end{split}
\end{equation} 
Hence, we have
 \begin{equation}
 X= \sqrt{2}(X_{\mu_{1}} + X_{\mu_{0}}) = x_{B} - x_{A}.
\end{equation}  

For $X=0$, both Alice and Bob have the same amount of wealth. Otherwise, for $X>0$ $(X<0)$, Bob has more (less) wealth than that of Alice. Thus, Charlie can conclude which millionaire has more assets and announce the result.
\end{description}

\subsection{Multiparty socialist millionaire problem \label{sec:SMP-n}}
We further discuss a more general scenario where $n$ parties wish to compare their assets \cite{liu2017efficient,ye2018multi}. It also serves our motivation that the CQD scheme proposed here has scalability for the purpose of its implementation for quantum networks.
With these intentions we have extended our SMP solution to the multiparty case, where the $n^{\rm th}$ party is requested by Charlie to encode $-(n-1)$ times his/her current wealth (in millions). 
The protocol is as follows.
\begin{description}
\item{Steps~1-5}. Same as Steps 1 to 5 of the previous protocol.
\item{Step~6}. Each of the following $n-1$ parties including Bob displace their transmitted mode by the same amount as their wealth.
\item{Step~7}. The $n^{\mathrm{ th}}$ millionaire displaces the mode by $-(n-1)$ times the amount of his/her wealth $x_{n}$.
\item{Step~8}. After checking for eavesdropping, Charlie performs Bell measurement on $a_{12}$ and $a_{61}$, to obtain    
     \begin{equation}
 X= \sqrt{2}( X_{\mu_{1}} + X_{\mu_{0}}) = \sum_{i=1}^{n-1}x_{i} - (n-1)x_{n}.
      \end{equation} 
For $X=0$,  Charlie concludes all millionaires have an equivalent amount of wealth. Otherwise, for $X>0$ or $X<0$, Charlie announces that the wealth of millionaires are not equal. 
\end{description}

Note that all the cases when $X\neq0$, the result is inconclusive. Thus, the present solution solves only a limiting case of the SMP known as quantum private comparison \cite{thapliyal2018orthogonal,ye2018multi}, where only equality of the assets of all the parties is to be verified. On the other hand, this also states that multiparty secure computation/communication becomes complex and thus the solution is not always trivially extendable.

\section{Security\label{sec:sec}}
Here we are going to discuss a set of individual attacks Eve can attempt and thereby we aim to establish that our protocol is secure against such attacks \cite{zhou2017new}. Generally, Eve may attempt the set of attacks discussed here in all three proposed schemes, unless stated otherwise; and thus security of all the proposed scheme is discussed together.
First of all, we will briefly discuss information leakage inherent in quantum dialogue based schemes (see \cite{banerjee2016asymmetric} and references therein for detail). As total information encoded in most of the quantum dialogue based schemes is twice the channel capacity $C$, such a scheme naturally leads to leakage of one-half of the encoded information as Eve's ignorance is equal to either Alice's or Bob's encoding. In the present case, Alice (or equivalently Bob) encodes a continuous message in both amplitude and phase quadratures; and thus unlike a DV QD, where Eve has to choose between only few possibilities (as 4 choices in \cite{nguyen2004quantum}), Eve has infinitely many options. Thus, Eve's ignorance is equal to the channel capacity $C$, which shows the advantage of CV over DV scheme, whereas information leakage--same as channel capacity--can be circumvented in CV CQD scheme by transmitting Charlie's random operations in a secure manner to Alice and Bob (as suggested in \cite{banerjee2016asymmetric}). 

\subsection{Mutual information of Alice and Bob}
The mutual information between Alice and Bob depends upon various parameters, such as transmittance (channel efficiency parameter) $\eta$, excess noise $\epsilon$, the intrinsic variance present in the beam $\sigma$, and the quantity $\Sigma_{ABC}$ which implies the variance introduced in the beam due to operations performed by Alice, Bob, and Charlie on the given mode. Mutual information between Alice and Bob in $X$ quadrature can be given as \cite{weedbrook2013continuous,usenko2011squeezed,weedbrook2010quantum,cerf2001quantum},
\begin{equation}
I^{X}_{AB}= log_{2}\bigg(1+\frac{\eta \Sigma_{ABC}}{\lambda + \eta(\sigma+\gamma-1)}\bigg).
\end{equation} 
Here, we consider a simpler case such that, $\eta=1$, $\epsilon = 0$,  $\lambda= 1$ (2) in case of homodyne (heterodyne) detection, and $\sigma^{2}=\frac{e^{-2r}}{4}$.  Further, we have $\Sigma_{ABC}=\Sigma_{A}+\Sigma_{B}+\Sigma_{S(C)}$, where $S(C)$ is the initial random displacements made by Charlie.
The measurement result obtained by Bob after homodyne detection yields
\begin{equation}
X=( \sqrt{2}(X_{\mu_{0}}) -x_{B} - x_{A} + x_{R_{B_{T_{4}}}} -x_{R_{A_{T_{4}}}})/\sqrt{2}
\end{equation}
corresponding to the amplitude quadrature,
where the information $x_{B}$ and $x_{A}$ is the encoded information while $x_{R_{B_{T_{4}}}}$ and $x_{R_{A_{T_{4}}}}$ are random operations publicly announced by the end of the protocol. Similarly, information is encoded on the phase quadrature as well. Therefore, total information exchanged during a run of the protocol is $x_{A}+x_{B}+y_{A}+y_{B}$, which would be $4N$ bits if discrete encoding is performed (say 12-bit using 3 bit encoding rule  in \cite{zhou2017new}). 
Charlie's random operations ${R_{A_{T_{4}}}}$ and ${R_{B_{T_{4}}}}$ (say $R_{C}$ collectively) serve analogous to the cryptographic switch \cite{srinatha2014quantum,thapliyal2015applications}, where Charlie can control the amount of information transferred between Alice and Bob to a continuously varying degree by only revealing part of $R_{C}$.    

\subsection{Disturbance Attack}
In this attack, Eve tries to mislead both parties either by sending random optical mode to either Bob (or equivalently Alice) or applying a random displacement operation while transmission. Note that Eve does not wish to extract the secret in this attack. This attack forbids the legitimate parties from accomplishing the task. To rescue the protocol from this particular attack, initially Charlie and Bob, and later Alice and Bob check for eavesdropping by looking for entanglement between their modes using the initial information; if they find at any stage, that the modes shared between both the parties are not entangled they discard the protocol.

{Specifically, the random displacement applied by Charlie and the initial public information lead to
\begin{eqnarray}
&&\ X_{a_{21}} = X_{a_{11}} + x_{R_{B}},\nonumber \\
&&\ X'_{\mu_{0}}  = (X_{a_{31}} - X_{a_{12}})/\sqrt{2}.
\end{eqnarray}
Since Bob measures $X_{a_{31}}$ and knows the initial information $X'_{\mu_{0}}$, he calculates the result to be announced by Charlie as $X= \sqrt{2}X'_{\mu_{0}}  - X_{a_{31}} $, and if Charlie announces the same measurement outcome then Bob can conclude that eavesdropping is not attempted. In a similar manner, one can verify in the case when Alice receives the mode from Charlie and when Alice sends her mode to Bob.}

\subsection{ Man in the middle attack}
Eve may also behave as an impostor and take on the identity of Alice (Bob) to Bob (Alice) and start exchanging information with the other party. An authenticated classical communication is used in the protocol which forbids Eve from this attack. 

{Eve may also intercept the optical mode sent from Charlie to Bob and sends an auxiliary mode to Bob. Though Eve is aware of the initial information $X_{\mu_{0}}$, it will not help her as both Bob and Charlie would announce measurement outcomes that would not be correlated. Specifically, Eve may prepare two-mode squeezed vacuum with same  $X_{\mu_{0}}$ and would attempt to escape undetected during Charlie to Bob (Alice) eavesdropping checking and subsequently intercept Alice to Bob transmission to decode Alice's secret. With the help of Bob's (Alice's) and Charlie's announcement of measurement outcomes, Eve may know the corresponding value of $x_{R_{i}}$ but the choice of $R_{A}$ and $R_{B}$ is random for each time frame. Therefore, Eve would not have any advantage.}

 \subsection{Cloning attack}
 Note that Charlie prepares two-mode squeezed vacuum state and sends both modes to Bob and Alice, respectively. Therefore, both modes are accessible to Eve at some instant of time (though not simultaneously). This may allow Eve to design a quantum cloning attack to make as exact copy as possible of Charlie to Alice transmitted mode and intercept Alice to Bob transmission to extract Alice's secret from that.  The  universal cloning machine \cite{scarani2005quantum} for $P \to Q$ in general acts in the following way
 \begin{equation}
\ket{\psi^{\otimes P}}\otimes \ket{r^{\otimes (Q-P)}}\otimes (\ket{A})\xrightarrow{U} \ket{\Phi}.
 \end{equation}
 Here, $\ket{\psi}$ is copied on the suitably chosen reference state $\ket{r}$ with an additional requirement of ancilla $\ket{A}$. Not only Eve is unable to clone it exactly, the random displacement operations by Charlie ($R_{C}$) forbid her to decode Alice's secret.
Further, the variance introduced by the cloned state in both quadratures corresponding to the transmitted mode would not be zero, i.e., the initial variance for the amplitude quadrature, say $\sigma_{C\rightarrow A}=\sigma+\Sigma_{S(C)}$, would change to $\sigma{'}_{C\rightarrow A} \neq \sigma_{C\rightarrow A}$ after being cloned by Eve. This would  leave detectable traces in entanglement checking.

\subsection{Beam splitter attack}

In this type of attack Eve uses a set of beam splitters to obtain information encoded by Alice, by applying a beam splitter in Charlie to Bob transmission and one during Alice to Bob transmission. In  first step she uses a beam splitter on the two modes, $a_{21}$ and $e_{11}$,  with a transmission coefficient $\beta_{1}$ resulting in
\begin{equation}
\begin{split}
&\ X_{a_{51}} = \sqrt{\beta_{1}}X_{a_{21}}  + \sqrt{1-\beta_{1}}X_{e_{11}},\\
&\ P_{a_{51}} = \sqrt{\beta_{1}}P_{a_{21}}  + \sqrt{1-\beta_{1}}P_{e_{11}},\\
&\ X_{e_{21}} = \sqrt{\beta_{1}}X_{e_{11}}  + \sqrt{1-\beta_{1}}X_{a_{21}},\\
&\ P_{e_{21}} = \sqrt{\beta_{1}}P_{e_{11}}  + \sqrt{1-\beta_{1}}P_{a_{21}}.\\
\end{split}
\end{equation}
Eve uses a quantum memory to store one of these modes $e_{21}$ and sends the mode $a_{51}$ to Bob via the quantum channel. In the second step,  Eve intercepts the transmitted mode $a_{42}$ while Alice to Bob transmission and uses her auxiliary mode $e_{12}$ as inputs of a beam splitter (with transmission coefficient $\beta_{2}$) giving the following results 
\begin{equation}
\begin{split}
&\ X_{a_{82}} = \sqrt{\beta_{2}}X_{a_{42}}  + \sqrt{1-\beta_{2}}X_{e_{12}},\\
&\ P_{a_{82}} = \sqrt{\beta_{2}}P_{a_{42}}  + \sqrt{1-\beta_{2}}P_{e_{12}},\\
&\ X_{e_{22}} = \sqrt{\beta_{2}}X_{e_{12}}  + \sqrt{1-\beta_{2}}X_{a_{42}},\\
&\ P_{e_{22}} = \sqrt{\beta_{2}}P_{e_{12}}  + \sqrt{1-\beta_{2}}P_{a_{42}}.\\
\end{split}
\end{equation}
Eve now sends the mode $a_{82}$ to Bob, while she keeps the mode $e_{22}$ with herself. Hereafter, she sends modes $e_{22}$ and  $e_{21}$ through two inputs of a beam splitter (with transmission coefficient $\beta_{3}$).  This operation can be defined as
\begin{equation}
\begin{split}
&\ X_{e_{32}} = \sqrt{\beta_{3}}X_{e_{22}}  + \sqrt{1-\beta_{3}}X_{e_{21}},\\
&\ P_{e_{32}} = \sqrt{\beta_{3}}P_{e_{22}}  + \sqrt{1-\beta_{3}}P_{e_{21}},\\
&\ X_{e_{31}} = \sqrt{\beta_{3}}X_{e_{21}}  + \sqrt{1-\beta_{3}}X_{e_{22}},\\
&\ P_{e_{31}} = \sqrt{\beta_{3}}P_{e_{21}}  + \sqrt{1-\beta_{3}}P_{e_{22}}.\\
\end{split}
\end{equation}
Subsequently, Eve chooses to measure the amplitude quadrature of mode $e_{32}$ and the phase quadrature of $e_{31}$ to obtain Bob's secret. This leads to the following result in case of amplitude quadrature
 \begin{eqnarray}
  X_{a_{110}} &=&\sqrt{\beta_{2}\beta_{3}}X_{e_{12}}  + \sqrt{(1-\beta_{2})\beta_{3}}X_{a_{12}} +  \sqrt{(1-\beta_{3})(1-\beta_{1})}X_{a_{11}}+ \sqrt{(1-\beta_{3})(\beta_{1})}X_{e_{11}}  \nonumber \\
 & -& \sqrt{(1-\beta_{3})(1-\beta_{1})}x_{R_{B}} + \sqrt{(\beta_{3})(1-\beta_{2})}x_{R_{A}}+\sqrt{(1-\beta_{2})\beta_{3}}x_{A}. 
  \end{eqnarray} 
  In order to obtain any useful information, such as $x_{A}$, Eve would require the values of variables $x_{R_{B}}$ and $x_{R_{A}}$. Irrespective of this Eve would be detected during eavesdropping checking performed by Charlie and Bob. 

\subsection{Trojan horse attack}
There are few attacks by an eavesdropper, which exploit the experimental limitations or imperfections of devices and are mostly implementation dependent. Security against such attacks cannot be established theoretically, but experimentally can be maintained by using several isolators \cite{huang2016long}. {In the present case, Eve may attempt to obtain Alice's information using a Trojan pulse sent with mode $a^{*}_{32}$ going from  Charlie to Alice and {filter the same pulse while Alice to Bob transmission (as shown in Fig. \ref{fig:Th}).} However, Alice can circumvent such attack \cite{ma2016quantum} by using frequency filters and additional equipment to detect Eve's Trojan pulse.
\begin{center}
\begin{figure}[h]
\centering
 \includegraphics[scale=0.3]{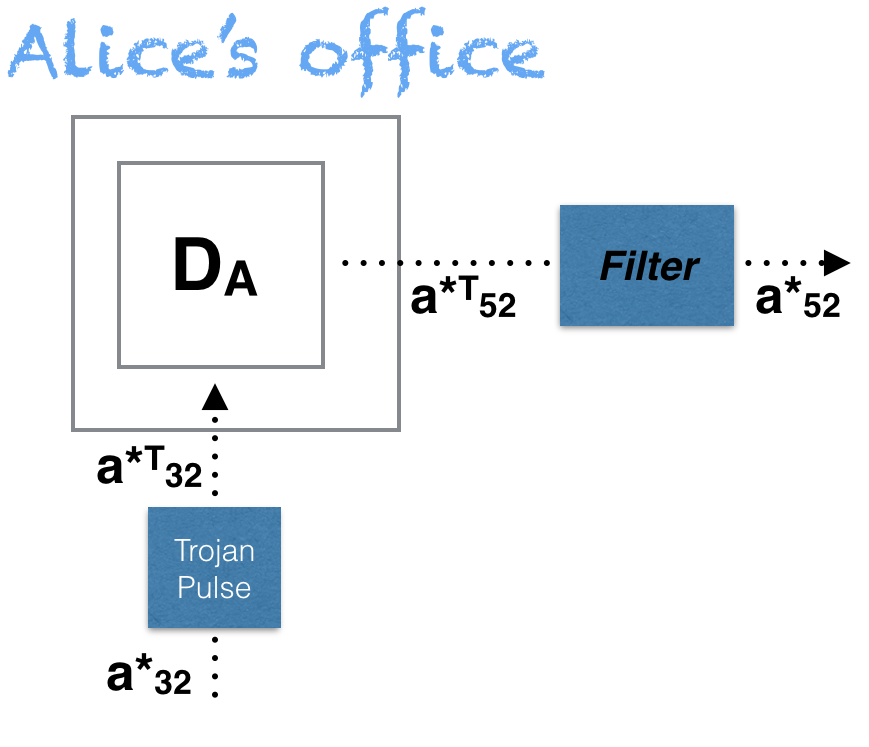}
\caption{\label{fig:Th}(Color online) A schematic diagram for Trojan horse attack.}
\end{figure}
\end{center}

\subsection{Malicious user's attack or internal attack}
A participant attack is difficult to circumvent than an outsider's attack. In these attacks, one of the parties would like to take advantage of the limited access to information to get the remaining secret. As an example, we assume that Charlie, who is the controller, makes a malicious attempt to obtain the information of Alice and/or Bob, either during Alice to Bob transmission or after Bob has announced the measurement result. Here, we show how our protocol is safe against such an attack, in which Charlie tries to intercept during Alice to Bob transmission and exploiting his power as a controller tries to deduce Bob's information. Specifically, Charlie knows the initial information and the random displacement he has applied on both the modes thus obtained, he tries to decipher the information encoded by Alice. 

Charlie may attempt to send an auxiliary mode to Bob and keep the mode entangled with Alice to perform Bell measurement and get her information. Interestingly, Charlie may even attempt to prepare a separable state. However, Alice and Bob perform entanglement checking in Step 5 of the proposed scheme, where these attacks by Charlie would be revealed.

Let us suppose Charlie tries to obtain Bob and/or Alice's information after Bob has announced the Bell measurement result, i.e.,
\begin{equation}
 X =  x_{R_{B_{T_{4}}}} - x_{R_{A_{T_{4}}}} -x_{B} - x_{A}.
\end{equation}  
Hence, Charlie would be able to obtain $
x_{B} + x_{A} =  x_{R_{B_{T_{4}}}} - x_{R_{A_{T_{4}}}} -X.
$
Though Charlie knows all the information on the right-hand side of the above equation, he would not be able to obtain any information hereafter unless he knows either the encoding by Alice or Bob. For this he may try to measure, clone, or guess Alice's mode, but that would reduce to an outsider's attack discussed previously.

\subsection{Malicious Alice/Bob in socialist millionaire problem}
So far we have discussed the attacks possible on all the schemes proposed here. Here, we are going to discuss some particular attacks relevant only in the solution proposed for SMP. Specifically, Alice and Bob wish to transmit their secrets to each other in CQD under the supervision of Charlie. In contrast, they do not trust each other in SMP and wish to take advantage of participant attack to know each other's encoding. Also, Charlie already knows the initial and final parameters of two-mode squeezed vacuum state prepared and measured by him. If somehow he gets to know Alice's (Bob's) encoding he would get the Bob's (Alice's) secret. Therefore, here Alice and Bob use a CV quantum key \cite{cerf2001quantum} to circumvent Charlie's participant attack. 

Further, Alice and Bob being distrustful parties may attempt to get inaccessible information of each other. First of all, one mode remains with Charlie and thus both Alice and Bob always encode and have access to a mixed state only. However, any attack by Alice (Bob) on Charlie-Bob (Alice-Charlie) transmission would reduce to an outsider's attack, which is already discussed. This includes attack by Alice where she would perform intercept and resend on Charlie-Bob transmission. Therefore, neither Alice nor Bob can take advantage of being a participant in the scheme to get each other's secret.  Further, in the multiparty version of SMP solution proposed here, collusion attacks by the participants can be circumvented in analogy of other circular quantum cryptographic schemes (\cite{sharma2017quantumauction,banerjee2018quantum} and references therein). 

\section{Conclusion\label{sec:con}}
In view of the advantages of CV communication schemes at metropolitan scale and controlled quantum communication in general, we have proposed here a new CQD protocol based on CV two-mode squeezed vacuum state. The scheme has several intrinsic advantages. For example,  it does not require tripartite entanglement unlike other controlled communication schemes; it can be used as a primitive to obtain solutions for several socioeconomic problems. As a particular example, we have discussed a CV solution of SMP using our proposed scheme. Due to use of only bipartite entanglement in the present scheme, it motivates us to look for a generalization of the present scheme which can be performed over quantum communication networks and would be implementable with the existing infrastructure. However, a close look at the multiparty SMP solution reveals that it only solves special case of the problem known as quantum private comparison, which can be attributed to the fact that with an increase in the number of parties,  complexity of the desired task also increases.

The security of the proposed scheme against some individual and participant attacks is also provided with the help of decoy pulses. The present scheme can be viewed as a CV counterpart of the idea of a quantum cryptographic switch which allows the supervisor to vary the information accessible to the receivers in a continuously varying degree. Further, the present scheme increases the ignorance of an eavesdropper as he needs to guess out of the infinitely many possible encoding operations by Alice (or equivalently Bob). Thus, it is less prone to the information leakage inherent with QD based schemes. 

The proposed scheme can also be used as primitive to design quantum cryptographic schemes for e-commerce and voting. We are further working on extension of such a protocol in case of $n$ parties and study the case where we might have more than one controller and how this may lead to a dynamic quantum communication network infrastructure. 

\textbf{ Acknowledgment:} AP and AS thank the Department of Science and Technology (DST), India, for support provided through the DST project No. EMR/2015/000393. KT acknowledges the financial support from the Operational Programme Research, Development and Education - European Regional Development Fund project no. CZ.02.1.01/0.0/0.0/16\_019/0000754 of the Ministry of Education, Youth and Sports of the Czech Republic.

\bibliographystyle{unsrt}
\bibliography{refs}
\end{document}